# Direct Imaging and Electronic Structure Modulation of Moiré Superlattices at the 2D/3D Interface

**Authors**
Kate Reidy[§,1], Georgios Varnavides[§,1,2,3], Joachim Dahl Thomsen[1], Abinash Kumar[1], Thang Pham[1], Arthur M. Blackburn[4], Polina Anikeeva[1,2], Prineha Narang[3], James M. LeBeau[1], Frances M. Ross[1*]

[1]Department of Materials Science and Engineering, Massachusetts Institute of Technology (MIT), Cambridge, MA, USA
[2]Research Laboratory of Electronics, Massachusetts Institute of Technology (MIT), Cambridge, MA, USA
[3]John A. Paulson School of Engineering and Applied Sciences, Harvard University, Cambridge, MA, USA
[4]Department of Physics and Astronomy, University of Victoria, Victoria, BC, Canada

[§]These authors contributed equally: Kate Reidy and Georgios Varnavides
*Corresponding author: fmross@mit.edu

**The atomic structure at the interface between two-dimensional (2D) and three-dimensional (3D) materials influences properties such as contact resistance, photo-response, and high-frequency electrical performance. Moiré engineering is yet to be utilized for tailoring this 2D/3D interface, despite its success in enabling correlated physics at 2D/2D interfaces. Using epitaxially aligned $MoS_2/Au\{111\}$ as a model system, we demonstrate the use of advanced scanning transmission electron microscopy (STEM) combined with a geometric convolution technique in imaging the crystallographic 32 Å moiré pattern at the 2D/3D interface. This moiré period is often hidden in conventional electron microscopy, where the Au structure is seen in projection. We show, via *ab initio* electronic structure calculations, that charge density is modulated according to the moiré period, illustrating the potential for (opto-)electronic moiré engineering at the 2D/3D interface. Our work presents a general pathway to directly image periodic modulation at interfaces using this combination of emerging microscopy techniques.**

## Introduction

Following the success of moiré engineering in modulating (opto-)electronic properties of graphene/hexagonal boron nitride (hBN) heterostructures[1,2] and twisted bilayer graphene[3–6], studies have extended the moiré toolbox to include systems such as double bilayer graphene[7], trilayer graphene[8], and van der Waals (vdW) heterostructures composed of transition metal dichalcogenides (TMDCs)[9,10] and hBN-graphene-hBN stacks[11,12]. Recently, moiré engineering has been extended beyond vdW heterostructures, to 3D/3D oxides[13]. Moiré engineering is yet to be utilized for tailoring the quasi-vdW interface between a 2D material and 3D metal. Engineering such 2D/3D interfaces is key to device applications where 2D materials make contact, through a well-controlled junction, to a 3D material such as a metal or semiconductor[14–16]. In contrast to 2D/2D heterostructures, moiré engineering at the 2D/3D interface requires the consideration of the stacking of atomic planes in the out-of-plane direction. 3D stacking introduces an additional tuning parameter in 2D/3D systems for modulating moiré properties that is not available in 2D/2D heterostructures[17].

The ability to image moiré superlattices directly is required to map electronic property modulation onto atomically-resolved structure[18]. Various techniques have been used to observe moiré superlattices. These include reciprocal space imaging *via* low energy electron diffraction (LEED)[19,20] and convergent beam electron diffraction (CBED)[21]; spatially resolved property measurement *via* scanning tunneling microscopy (STM)[22,23], atomic force microscopy (AFM) modalities[2,24], near-field optical microscopy[13], and infrared nano-imaging[5]; and imaging of transmitted intensity *via* high-resolution and dark field (scanning) transmission electron microscopy, (S)TEM[25,26]. Of these techniques, STM and (S)TEM are the only two that exhibit real-space atomic resolution. STM is widely used to characterize moiré patterns in 2D materials on bulk substrates, such as graphene on Ru[27], Ir[28], and Cu[29]. However, STM measurements are challenging for deeply buried interfaces and for the suspended layers that are gaining traction in 2D device physics[30,31]. (S)TEM, on the other hand, provides detailed information for suspended moiré systems fabricated from solely 2D materials[25,26]. Interpretation is more challenging for 2D/3D interfaces due to the necessity of considering the 3D structure of layers away from the interface[32,33]. This has resulted in discrepancies in periodicity measurement between imaging techniques[22,33]. The $MoS_2$/Au{111} system highlights these challenges, with different values reported for the periodicities of superlattices measured via STM and (S)TEM, 32 Å and 18 Å respectively[22,33].

To reconcile such discrepancies and map moiré structure-property relations at the 2D/3D interface, we combine an analytic convolution technique and a range of STEM imaging techniques (integrated differential phase contrast (iDPC) and four-dimensional (4D) STEM) to decouple the spectrum of higher order moiré patterns. We investigate $MoS_2$/Au{111} as a model 2D/3D system, relevant to TMDC (opto-) electronics[14], and also examine hBN/Au{111} relevant in plasmonics[46,47]. iDPC STEM images the phase of the sample transmission function, enabling direct interpretation as the projected electrostatic potential in thin samples[34–36]. 4D STEM is a rapidly developing technique in which a pixelated array detector is used to collect a convergent beam electron diffraction pattern at each probe position in the STEM image. The resulting 4D dataset can be filtered post-acquisition to produce reconstructions such as bright field, annular bright field, annular dark field (ADF), ptychography, and iDPC[37]. 4D STEM has been applied to materials including Cu[38], $ZrO_2$[39], $LiNiO_2$[40], $DyScO_3$[41], graphene[42], $MoS_2$[43] and $WS_2$[44], with 2D materials particularly well-suited due to their thinness[45]. We show that iDPC and 4D STEM are able to decouple higher order moiré periods to form real space images of the moiré pattern at the 2D/3D interface of $MoS_2$/Au{111}, revealing the crystallographic 32 Å period and explaining the difference compared to conventional (S)TEM in terms of projection effects of the ABC stacking of the 3D metal. We then use *ab initio* electronic structure calculations to corroborate that $MoS_2$/Au{111} charge density modulation is concentrated at the interface and follows the 32 Å moiré periodicity. Together these findings demonstrate the utility of direct imaging *via* iDPC and 4D STEM for understanding the structure and electronic properties of 2D/3D heterostructures.

## Results

**Microscopy of $MoS_2$/Au{111} system.** An example of the $MoS_2$/Au{111} interface is shown in Figure 1. In contrast to the mechanical transfer processes employed for fabricating vdW heterostructures, the 2D/3D systems studied here were formed by direct epitaxial growth[48] in ultra-high vacuum conditions (Methods). The resulting samples consist of flat, faceted Au{111} nanoislands with an average edge length of 25 nm and height of 8 nm (Figure 1a, Supplementary Figure 1) that are epitaxially aligned on suspended $MoS_2${0001} (Figure 1b), with uniform moiré periodicities across micrometre-scale areas. Selected-area electron diffraction (SAED) confirms 0º rotation between Au and $MoS_2$ with a standard deviation of 0.2º (Supplementary Figure 2). In Figure 1b and other SAEDs, we observe spots indexed as $1/3\{422\}$ Au reflections. These are classically forbidden for the FCC structure but their presence is consistent with Au nanoisland literature[49] (Supplementary Note 1). High resolution (HR) TEM shows that the islands are single crystalline, with no evidence of misfit dislocations and low angle grain boundaries (Figure 1c). The discontinuity in the moiré pattern visible at some boundaries arises from island coalescence. Here, both rigid body displacements and twin boundaries arise from stacking faults between coalesced islands (Figure 1d, blue arrows).

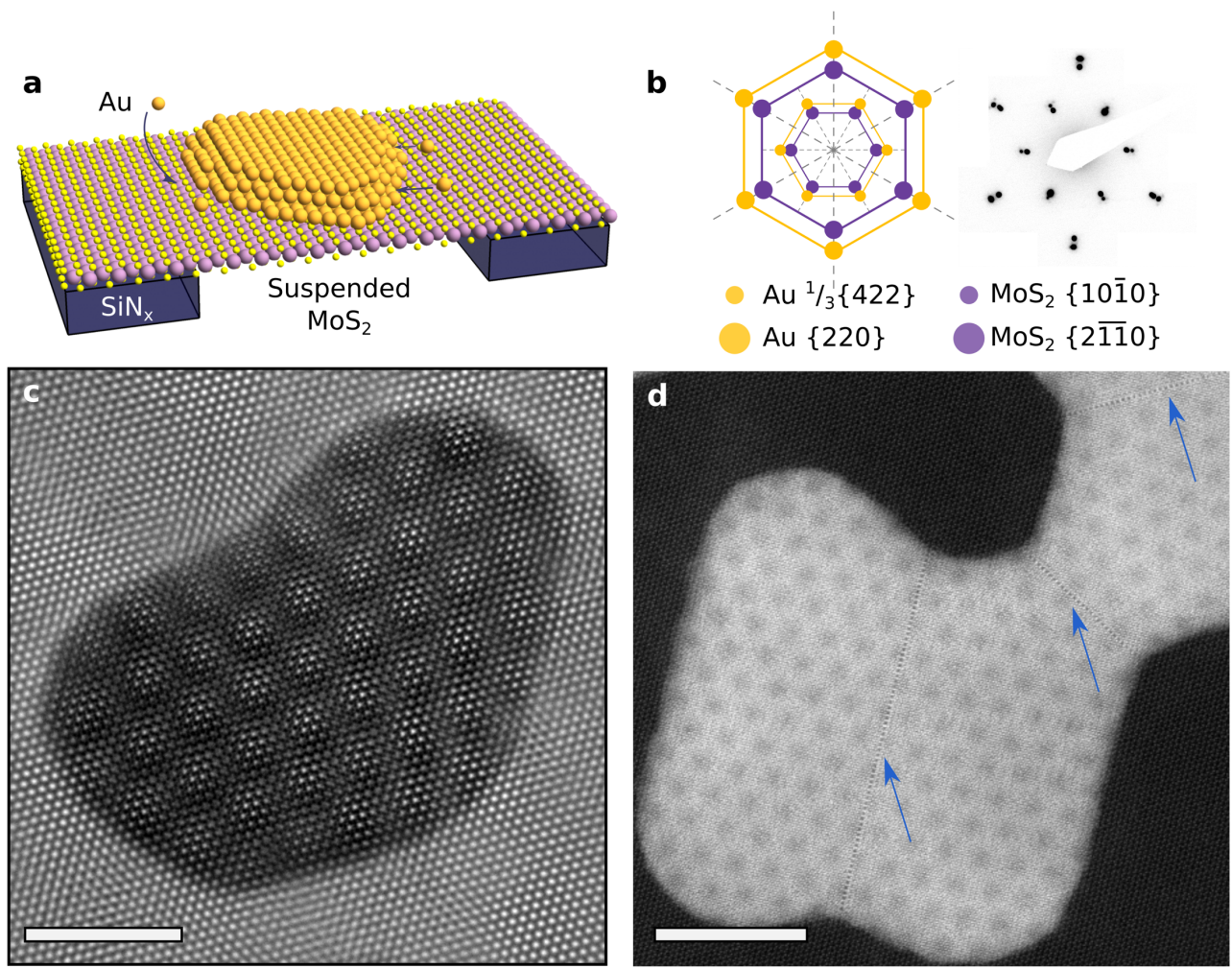

**Fig. 1 HRTEM and STEM demonstrating epitaxial MoS$_2$/Au{111} moiré. a,** Schematic of epitaxially aligned Au deposited on suspended MoS$_2$ on a SiN$_x$ TEM grid. Orange atoms represent Au, yellow S, and purple Mo. SiN$_x$ membrane is shown in dark blue. **b,** Reciprocal space model and experimental selected area electron diffraction pattern of the Au {111} zone aligned on MoS$_2$ {0001}, with weak intensity $1/3\{422\}_{Au}$ spots (see text) aligned with $\{10\bar{1}0\}_{MoS2}$ and higher intensity $\{220\}_{Au}$ spots aligned with $\{2\bar{1}\bar{1}0\}_{MoS2}$. Orange dots represent frequencies from Au crystal planes, while purple represent frequencies from MoS$_2$ crystal planes. **c,** HRTEM image of Au nanoisland on MoS$_2$, showing apparent 18 Å-period moiré pattern. Scale bar, 40 Å. **d,** High angle annular dark field (HAADF) STEM image. Scale bar, 80 Å. Coalescence boundaries are marked by blue arrows.

The uniform moiré periodicity and sinusoidal intensity modulation show that the Au and MoS$_2$ lattices are undistorted in the plane of the interface, even near island edges. This is different from the case of twisted vdW structures, which frequently display reconstructions[25,26]. The absence of distortion can likely be attributed to weak quasi-vdW bonding at the MoS$_2$/Au{111} interface[14]. Motion of islands at room temperature is consistent in suggesting loose binding of the Au nanoislands to the underlying substrate (Supplementary Movie 1). During their motion, the islands exhibit rotation up to 0.3º, visually amplified in the angle of the moiré pattern (Supplementary Figure 2). These MoS$_2$/Au{111} interface characteristics are consistent for a range of Au thicknesses and uniform across samples (Supplementary Figure 3).

**Moiré site inequivalence.** At first glance, the period of the MoS$_2$/Au{111} moiré superlattice in Figure 1c,d is 18 Å. While this is in agreement with previous HRTEM studies[33], it is a consequence of the projective nature of conventional (S)TEM imaging. To illustrate this, one can consider a thought-experiment in which the out-of-plane coordinate of the 3D Au{111} structure is ignored; this results in a "projected" hexagonal Au lattice with atomic spacing of 1.66 Å, which indeed yields a moiré pattern of 18 Å with the MoS$_2$ substrate. A more accurate view of electron scattering through the Au crystal requires us to include the full face-centred cubic (FCC) Au structure, as shown in Figure 2a,b. Consider a location where an Au atom from the A layer (orange) is directly above a pair of S atoms, as in the centre of Figure 2c-top. This site repeats every 32 Å, shown by the orange squares in Figure 2b. Sites that appear similar (red and blue squares in Figure 2b) instead have Au atoms from the B or C layers above the S atoms (Figure 2c-middle, 2c-bottom). The inequivalence of the three sites can be further illustrated via radial distribution functions (RDFs), which show the quantitative difference in atomic locations (Figure 2d). Although HRTEM (Figure 1c) and STEM (Figure 1d, 2e,g) do not distinguish the three sites, we find that iDPC STEM imaging (Figure 2f), sensitive to the projected electrostatic potential[50,51], shows small changes in contrast that are statistically significant (Figure 2h, Methods) and are confirmed by multislice simulations (Supplementary Figure 4). iDPC can therefore detect the true 32 Å moiré cell at the MoS$_2$/Au{111} interface. However, although this modulation is qualitatively and statistically observable, the translation and rotation of the quasi-vdW islands, as in Supplementary Movie 1, preclude a quantitative analysis.

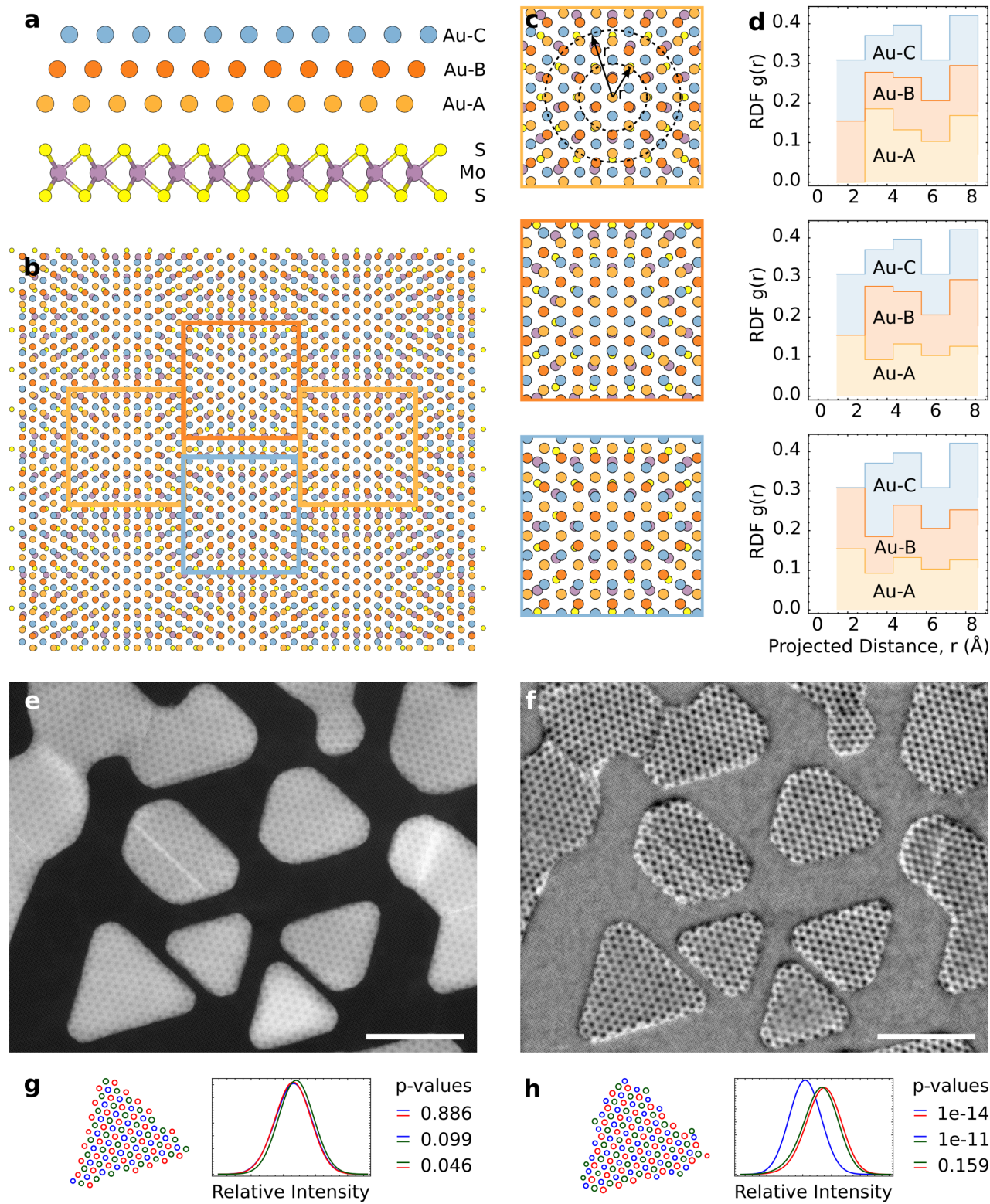

**Fig. 2 Atomic models, RDFs, HAADF, and iDPC characterization of 32 Å moiré structure. a,** Atomic model [100] zone axis for the 32 Å moiré. Note the Au atoms have been coloured differently (orange, red, blue) to highlight relative stacking of A, B, and C sites. **b,** Plan view atomic model for the 32 Å moiré. Boxed areas represent three inequivalent sites in the 32 Å moiré. **c,** Close up plan-view image of each of the sites highlighted in **b.** $r$ is the projected distance from the central aligned sites, and the dotted circles show two representative $r$ values. **d,** Corresponding RDFs of the three inequivalent sites in the 32 Å moiré. **e,** HAADF and **f** iDPC STEM images showing the (apparent) 18 Å and 32 Å moiré cells, respectively. Scale bars 200 Å. **g, h,** Relative intensity distributions and statistical variation of the three inequivalent sites in the corresponding images. The equivalent disk radius for each spot was calculated and partitioned to inequivalent sites (red,green,blue). The histograms were smoothed using a gaussian kernel of radius 0.5 Å for visual clarity.

Although visible, the iDPC signal from the 32 Å moiré is weak. To consider the full set of spatial frequencies of the Au{111} FCC crystal and obtain a clear real-space image of the 32 Å moiré, we turn to a reciprocal space convolution theorem to predict the entire spectrum of possible moirés in the $MoS_2$/Au{111} system (Methods). The geometric interpretation of the convolution theorem indicates that periodicities arise from the pairwise vectors connecting all spatial frequencies of the $MoS_2$ and Au lattices[52] (Figure 3a). In Supplementary Figure 5 and Supplementary Table 1, we calculate these periodicities and intensities as a function of rotation angle between the two crystals. The four largest periodicities are shown in Figure 3b. Additional higher order moirés are also predicted which often exhibit smaller periodicities and weaker intensities (Supplementary Figure 5). At zero rotation, we indeed recover the 32 Å moiré period, alongside the apparent 18 Å moiré (Figure 3b). Note that 32 Å moiré periodicity is obscured by the 230% higher intensity reflections of the 18 Å period convolution. We confirm this assignment of moiré periods by showing the experimental fast fourier transform (FFT) of Figure 1c (Methods, Figure 3c). The moiré superlattice periods emerge as two sets of satellite peaks around the central beam spot. The simulated diffraction pattern in Figure 3d is in quantitative agreement with the FFT of the acquired image (Figure 3c), predicting all higher order moiré periodicities at the 2D/3D interface.

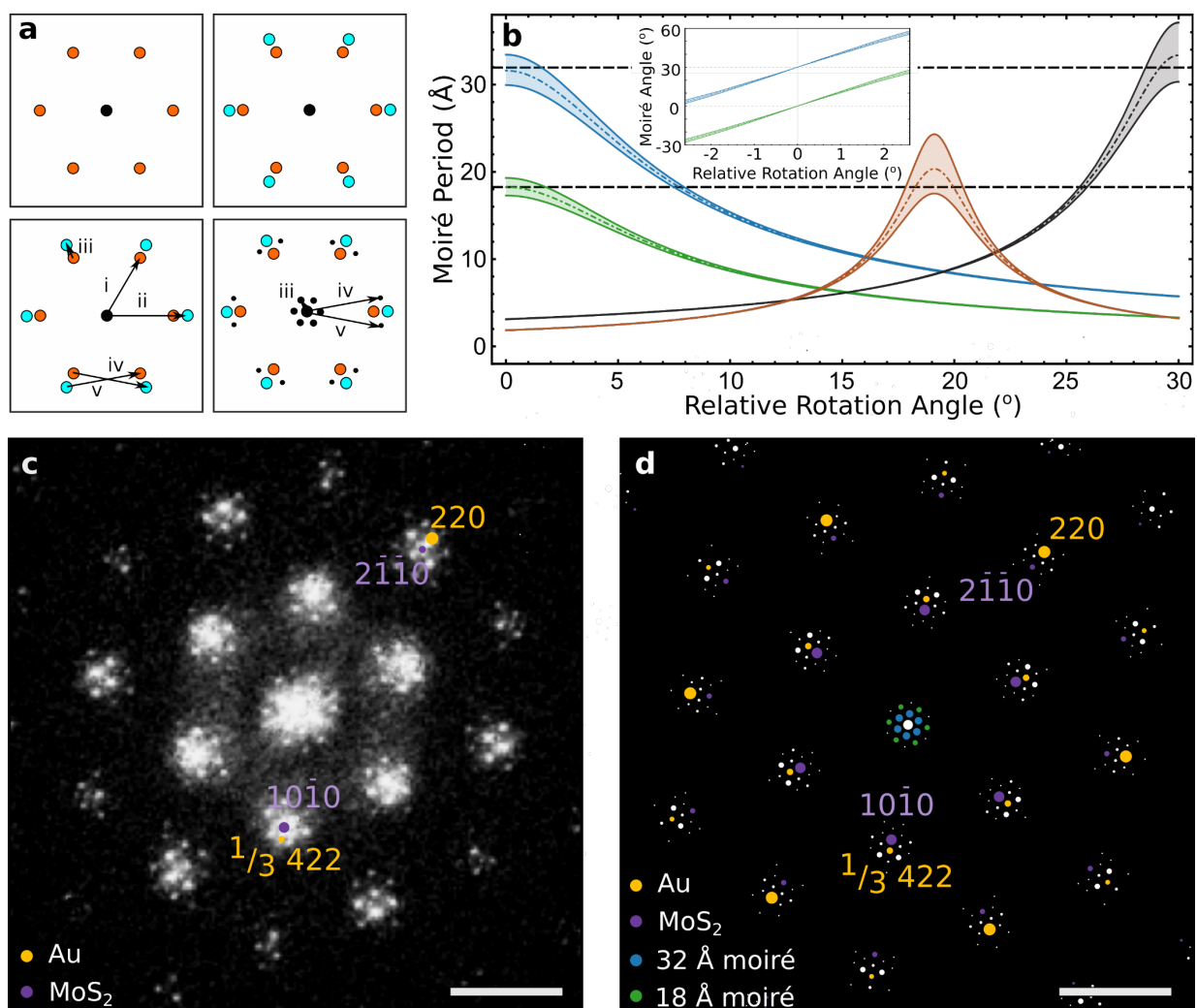

**Fig. 3 Geometric convolution technique to predict moiré spectrum. a,** Schematic representation of satellite spot generation. Spatial frequencies due to a single lattice shown in the top left panel (orange) are overlaid on those arising from a second lattice on the top right (cyan). The convolution of these two sets of spatial frequencies (i,ii) can be understood as the pairwise vectors connecting spatial frequencies of the two lattices (iii,iv,v - bottom, left). These convolutions generate moiré frequencies (iii, iv, v) shown as black dots in the bottom right panel. **b,** Calculated moiré period vs rotation angle for the four largest moiré supercells in the MoS$_2$/Au{111} system, illustrated for small (± 1 %) Au lattice strain. Dot dashed lines represent 0% strain, while the two solid lines on either side represent ± 1 % strain as a bound. Black dashed lines represent the experimentally observed moiré periods from the FFT, two of which (18 Å and 32 Å) are predicted at 0º relative rotation angle. The moirés are colour coded according to the reflections they arise from, with blue arising from the $\{2\bar{1}\bar{1}0\}_{MoS2} : \{220\}_{Au}$, green $\{10\bar{1}0\}_{MoS2} : {}^{1}/_{3}\{422\}_{Au}$, orange $\{6\bar{3}\bar{3}0\}_{MoS2} : \{642\}_{Au}$, and grey $\{20\bar{2}0\}_{MoS2} : \{220\}_{Au}$ reflections respectively. The inset shows the variation of moiré angle with relative rotation angle near 0º. **c,** FFT of atomic resolution HRTEM image of the MoS$_2$/Au{111} image in Figure 1c showing ${}^{1}/_{3}\{422\}$ reflection and two visible moiré periodicities around central spot. Illustrative orange dots represent frequencies from Au crystal planes, while purple represent frequencies from MoS$_2$ crystal planes. Scale bar, 0.5 Å$^{-1}$. **d,** Simulated FFT for Au/MoS$_2$ generated via the geometric convolution technique with each spot

coloured to show its origin (orange: Au, purple: MoS$_2$, blue: 32 Å crystallographic moiré, green: apparent 18 Å moiré). Area of spots is proportional to absolute intensity, but with inner moiré spots magnified 2x for clarity. Scale bar, 0.5 Å$^{-1}$.

To extract a real space image of the weak 32 Å moiré, we employ the technique of 4D STEM (Figure 4a)[37]. Subsequently we select, with a virtual annular dark field (ADF) detector, an annular area of each diffraction pattern to reconstruct an image from the average pattern (Figure 4b) using certain diffraction spots only. Using an annulus that includes the Au{220} spots and the MoS$_2$ {2$\bar{1}\bar{1}$0} spots (Figure 4c, Methods), we observe the high intensity 18 Å moiré pattern (Figure 4e). The moiré shows uniform periodicity and sinusoidal intensity modulation across the islands. The symmetry is reduced to periodic line patterns in some areas due to sample tilt, but 18 Å periodicity appears across all islands. If instead we generate a second virtual ADF image using the weaker ⅓{422} Au and {10$\bar{1}$0} MoS$_2$ reflections (Figure 4d), we observe a hexagonal pattern of spots with 32 Å moiré periodicity, consistent with our predictions from geometric convolution and the true crystallographic moiré accounting for 3D structure (Figure 4f).

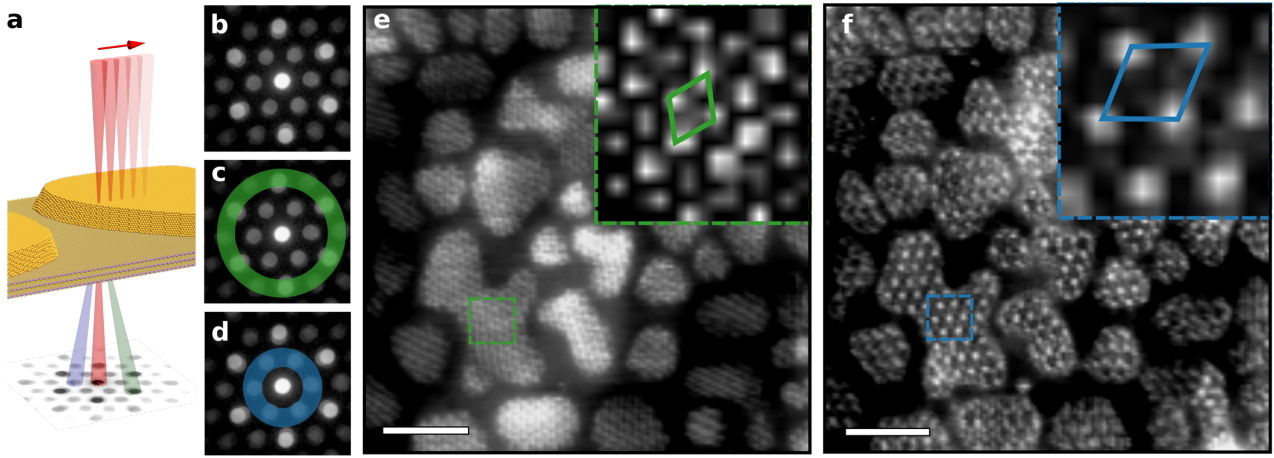

**Fig. 4 4D STEM imaging of 18 Å and 32 Å moiré periodicities. a,** Schematic of 4D STEM technique showing rastered beam (red) on MoS$_2$/Au{111} with corresponding CBED pattern at each point. The green and blue scattered beams are centred on the spots of their respective annuli shown in c and d. **b,** CBED pattern formed by averaging patterns collected over the entire scan area, **c,** 4D STEM annulus used to isolate 18 Å moiré periodicity (angular range 31-43 mrad - green) and **d,** 4D STEM annulus used to isolate 32 Å periodicity (angular range 11-24 mrad - blue). **e,f** Virtual ADF STEM images revealing 18 (green) and 32 Å (blue) period moirés, respectively. Scale bar, 200 Å. Insets show unit cells.

**Charge density modulation.** To explore the impact of the moiré periodicity on ground state charge density of our 2D/3D structure, we next turn to *ab initio* electronic structure calculations (Methods). Figure 5a shows a calculated isosurface of ground-state charge density difference for MoS$_2$/Au{111} in side-view. The charge density difference is concentrated at the interface, specifically on the upper S layer of atoms, with some penetration to the underlying Mo layer. On the Au side, the charge density difference is concentrated on the first atomic plane, with negligible charge density found in the second Au{111} layer. The charge density modulation due to the 2D/3D interface indeed has a periodicity of 32 Å (Figure 5b). To quantify the effect of moiré modulation on band structure and density of states, the supercell electronic states can be unfolded onto a single MoS$_2$ unit cell (Figure 5c). Accounting for the 32 Å moiré, band structure calculations are in agreement with prior angle resolved photoemission spectroscopy and scanning tunnelling spectroscopy measurements of the MoS$_2$/Au{111} system[53].

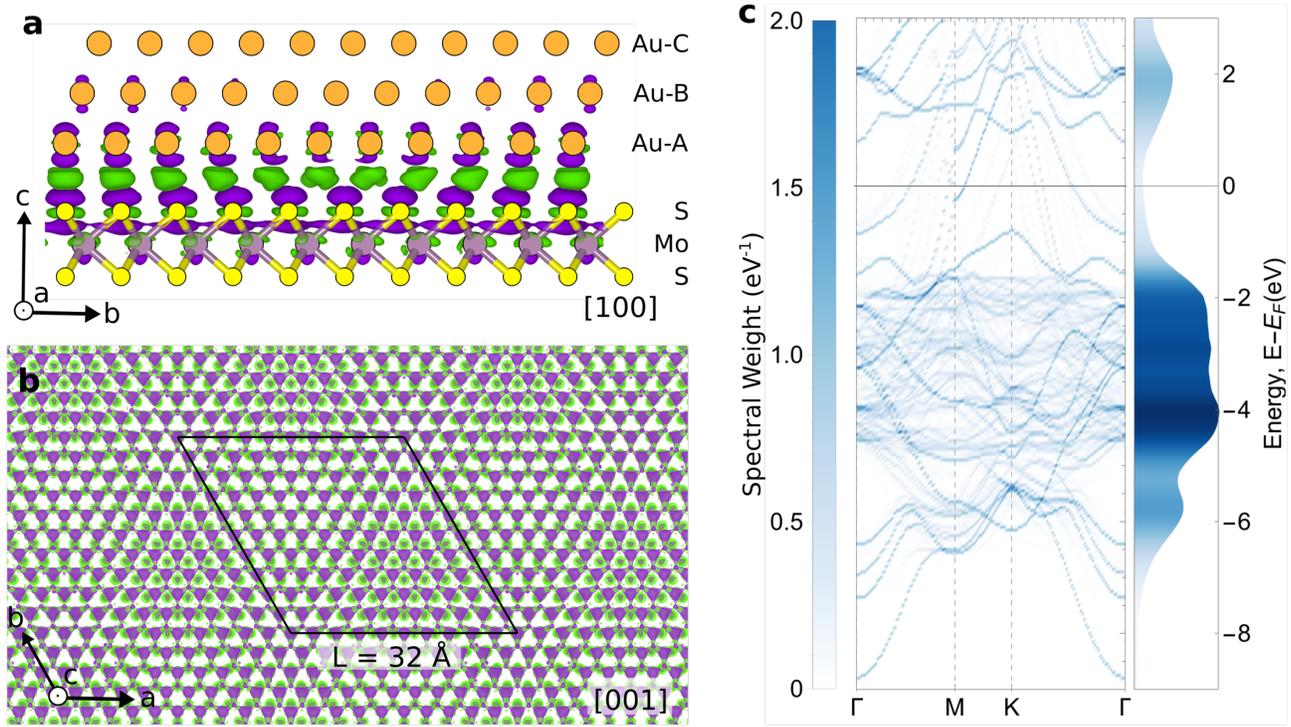

**Fig. 5 Electronic structure calculations at the 2D/3D interface a,** Charge density difference viewed down a [100] cross section of the 32 Å commensurate moiré ($11_{Au}$ x $10_{MoS2}$ superstructure). Orange atoms represent Au, yellow S, and purple Mo. Purple denotes negative and green positive charge density isosurface contours. **b,** Calculated charge density difference at the MoS$_2$/Au{111} interface, as viewed down the [001] axis, showing electronic modulation following the 32 Å moiré periodicity. Black line indicates the 32 Å crystallographic moiré unit cell. **c,** Unfolded band structure for MoS$_2$/Au{111} system (left) and corresponding density of states (right). Colour corresponds to the band's spectral weight.

**Application of the method to the hBN/Au{111} interface.** To explore the generality of the geometric convolution technique, we also apply it to the hBN/Au{111} structure (Supplementary Figure 6). Here, the Au lattice is rotated by 10º with respect to the hBN, leading to a more complex situation than the symmetric 0º epitaxy of MoS$_2$/Au{111}. This 10º rotation leads to a strong moiré periodicity of 11 Å. Prediction *via* geometric convolution technique is necessary in these rotated systems to uncover higher order moirés. For this interface, the convolution technique predicts and explains an additional 19 Å periodicity observed experimentally and re-creates the experimental diffraction pattern (Supplementary Figure 6). The predictions of the convolution technique rely solely on inputs of crystal structure, lattice parameters, and rotation. This analysis, as well as previous literature, illustrate the wide applicability of the technique in moiré analysis[52].

## Discussion

For the $MoS_2/Au\{111\}$ interface, the combination of several different imaging techniques with electronic calculations provides a clearer picture of the moiré structure than is possible with any single measurement. In HRTEM the 18 Å moiré is the strongest visible, leading to the possibility of erroneously predicting that electronic properties should be modulated with this period. Our calculations reveal that electronic modulation instead follows the true crystallographic 32 Å periodicity. This periodicity is hidden in conventional TEM due to projection effects. Instead, for this interface, 4D STEM imaging, combined with a geometric convolution analysis of the full moiré spectrum, allows a direct real-space observational link between atomic structure and moiré-induced electronic modulation at this 2D/3D interface. The combination of analysis techniques also explains the discrepancy between moiré patterns observed by TEM and STM at this 2D/3D interface. These results highlight electronic modulation at the 2D/3D interface, as well as showcasing the growing opportunities for advanced STEM techniques for direct imaging of moiré structures at the atomic scale.

We envision the coupled application of 4D STEM and geometric convolution theorem presented here for analysis of the 2D/3D interface could also be extended to the direct imaging of higher order moirés in systems with multiple interfaces and could expand opportunities across the field of moiré engineering. Pertinent examples of its potential application lie in multiple overlaid moiré superlattices, which have been found to coexist in stacked vdW heterostructures such as hBN-graphene-hBN stacks[11,12], or in so-called "moiré of moirés" structures from relaxation of twisted trilayer graphene and $WSe_2$[54]. Although the effects of these coexisting moirés have been reported, they have not yet been directly imaged. This is because the overall moiré observed in HRTEM and STEM is a convolved projection of all the moirés in the system. Using 4D STEM and geometric convolution, moiré characterisation could in theory be performed at each interface in the structure by highlighting the relevant diffraction spots. Virtual ADF images could then be used to decouple and directly image each separate moiré. Moreover, 4D STEM could enable simultaneous mapping of crystal orientation, strain, sample thickness, polarization, electric fields, and 3D ptychographic reconstructions of relevant moiré structures[37]. To date, most 2D/3D moiré investigations (including this study) have focussed on epitaxially grown interfaces exhibiting a single orientation. However, future practical development of 2D/3D moiré engineering will require complete control of the structure and orientation of 2D and 3D materials. Emerging fabrication methods using direct transfer of a 3D metal, such as Au, onto 2D materials[55], or nanomechanical rotation of a 3D nanocrystal using AFM or STM cantilevers[56] suggest that such control of interfacial orientation is increasingly feasible, extending opportunities of 2D/3D moiré engineering.

## Methods

**Suspended $MoS_2$ sample fabrication.** Custom TEM chips were fabricated, consisting of a $SiN_x$ membrane supported on Si, with 9 holes each 4 μm in diameter. We employed a wedging transfer process to suspend $MoS_2$ on these TEM grids[57]. Thermally grown 90 nm $SiO_2$/Si wafers were pre-treated with oxygen plasma and $MoS_2$ was mechanically exfoliated onto them using conventional Scotch tape method. Flakes of suitable thickness were identified by their contrast in optical microscopy. A solution of 25 g cellulose acetate butyrate (CAB) in 100 ml ethyl acetate was spin coated onto the sample and baked at 80 °C for 6 minutes. $MoS_2$ flakes were transferred to perforated silicon nitride ($SiN_x$) TEM grids using a wedging transfer technique[57]. Here, a scalpel is used to cut the CAB around the desired flake. A drop of deionised water can then be intercalated between the CAB and $SiO_2$/Si surface and the entire flake transferred to the TEM grid with the CAB polymer handle using a tweezers.

The transferred flakes were baked at 140 ºC for 5-10 minutes to improve adhesion. After dissolving the CAB in acetone for 15 mins, the flakes were dipped in isopropanol and dried using a critical point dryer.

**Ultra-high vacuum (UHV) epitaxial deposition.** To create epitaxial nanoislands, UHV deposition is used. This reduces impurities trapped at the metal-2D interface[58]. The main source of interfacial impurities is polymer residues, which create heterogeneous nucleation sites. Therefore, polymer residue remaining on the 2D material nucleates non-epitaxially aligned nanoislands (Supplementary Figure 1). This combination of CAB polymer and heat treatment is effective in removing carbon and polymeric contamination[31]; material transferred using other polymers such as PMMA cannot be cleaned as effectively. $MoS_2/SiN_x$ substrates were loaded into UHV sample preparation chambers and cleaned of residual polymer by heating resistively in UHV to ~550°C for several hours. Au deposition was carried out in the same multichamber UHV system (base pressure $2 \times 10^{-9}$ Torr), and was deposited from a homebuilt K-cells using sheet metal placed in a BN crucible at a rate of 0.5 Å/min. The deposited thickness was calibrated by measuring the evaporation rate with a quartz crystal microbalance immediately before and after deposition. AFM analysis of island thickness were performed in a Veeco Metrology Nanoscope V in tapping mode. There is no intentional heating during deposition, but thermocouple measurements show that the sample temperature rises to 50-60 °C.

**TEM imaging and data analysis**. A field-emission TEM (JEOL 2010F) was used for selected area electron diffraction and bright-field imaging, operated at 200 kV. HRTEM imaging was performed with a Hitachi HF-3300V with CEOS BCOR imaging aberration corrector, operated at 60 kV. Figure 1c was obtained from a drift corrected mean of 25 images, where each image was an 8 second exposure, so the total exposure on to the camera was 200 seconds. The electron flux was $500 e^-/Å^2/sec$ so the final image exhibited a total ~100,000 $e^-/Å^2$. Drift tracking over the images gave an average drift of < 7 pm/sec, although most images exhibited less drift. FFTs and line-scans were obtained using Fiji ImageJ software. FFTs of the real-space image are used for observing moiré peaks instead of SAED patterns since moiré peaks cannot be observed using at the energies (80−300 keV) used in TEM[59]. The FFT in Figure 3c was produced by multiplying the source image (Figure 1c) by a Hanning window prior to taking its FFT to minimise streaking incurred due the to 'hard edge' of the discrete bound of the image.

**Multislice Image Simulations.** STEM image simulations in Supplementary Figure 4 were performed using an orthorhombic supercell consisting of 3 Au layers on an $MoS_2$ monolayer (7956 atoms), sliced along the [001] direction. A repeating unit from the supercell was cropped and simulated using a custom Python-based STEM image Simulation software. Simulation parameters similar to experiments is used, accelerating voltage- 60 kV, convergence angle-24.7 mrad, collection angle: 25-153 mrad (ADF) and 6-24 (iDPC). Simulated ADF and iDPC images were convolved with a gaussian kernel having FWHM as 80 pm, approximately accounting for the finite effective source size.

**4D STEM imaging and data analysis.** 4D STEM imaging was performed with a probe-corrected Thermo Fisher Scientific Themis Z G3 60-300 kV S/TEM operated at 60 kV with a beam current of 50-60 pA in the microprobe mode and a semi-convergence angle of 5.42 mrad using an Electron Microscopy Pixel Array Detector. The equivalent probe size used in Figure 4 was ~1 nm and the pixel size was 0.813 nm. Virtual ADF STEM images were generated from 4D STEM dataset using virtual detectors using the '4D STEM Explorer' program.[60] The HAADF and iDPC images in Figure 2 were acquired at 200 kV, 25 mrad convergence angle, and a current of 30 pA. The quantification in Figure 2g,h was performed as follows: Input images were convolved with the Laplacian of a gaussian kernel with radius 3.75 Å prior to peak detection. Peaks were segmented using a watershed transform and an equivalent disk radius for each spot was calculated and partitioned to inequivalent sites (red,green,blue). A Student t-test was used to test the null hypothesis that the different sample means were equal, at the 0.001 significance level.

**Geometric convolution technique.** The geometric convolution code was implemented in the computational package Wolfram Mathematica 12.0 and builds on a model previously described for hexagonal lattices[52]. Frequencies arising from the superposition of the two lattice functions were obtained by the convolution theorem, F{t x b} = F{t} F{b}, where t and b are the top and bottom lattice functions respectively, F{} denotes the Fourier transform, and $\otimes$ denotes the convolution operation. All possible spatial frequencies arising from observed spots in the SAED/FFT were initially obtained and we make no assumptions in the simulation other than the bulk structure of Au and {111} orientation. The full set of spatial frequencies of the FCC crystal along the [111] zone axis were used to calculate the moiré periods for all possibilities within the experimentally observed FFT as a function of the relative rotation while allowing for small (± 1 %) Au lattice strain. The angles can also be calculated (Supplementary Figure 2). We then evaluate the most likely candidates to explain the experimentally measured moiré periods and angles (Supplementary Figure 5).

**Electronic structure calculations.** The ground-state charge density difference ($\Delta\rho$) between the Au/MoS$_2$ heterostructure ($\rho_{Au/MoS_2}$), and pristine Au ($\rho_{Au}$) and MoS$_2$ ($\rho_{MoS_2}$) is given by

$$\Delta\rho = \rho_{Au/MoS_2} - \rho_{Au} - \rho_{MoS_2} \qquad (1)$$

Density functional theory calculations were carried out using the projector augmented wave method implemented in the Vienna *ab initio* simulation package, VASP[61,62]. We account for the vdW dispersion interactions using the generalized gradient optB86b-vdW functional[63]. We use a cut-off energy of 400eV on an equivalent Monkhorst-Pack k-points grid of 40x40x1 MoS$_2$ unit cell (and similar density supercell). Bandstructure unfolding was performed using the BandUP code[64].

## Data Availability
The authors declare that the main data supporting the findings of this study are available within the article and its Supplementary Information files.

## Code Availability
Code available upon request from the authors.

## Acknowledgements

This work was carried out with the use of facilities and instrumentation supported by NSF through the Massachusetts Institute of Technology Materials Research Science and Engineering Center DMR - 1419807. This work was carried out in part through the use of MIT.nano's facilities. G.V. and P.N. acknowledge funding from the NSF Award DMR-1905295 and from the Office of Naval Research (ONR) Grant Number N00014-18-1-2691 for the theory and computation in this work. P.N. is a Moore Inventor Fellow supported through Grant GBMF8048 from the Gordon and Betty Moore Foundation. K.R. acknowledges funding from funding a MIT MathWorks Engineering Fellowship and an OGE MIT Fellowship. J.D.T acknowledges support from Independent Research Fund Denmark though grant number 9035-00006B. The authors would like acknowledge Michael Tarkanian for help in manufacturing TEM sample holders; Professor Pierre Stadelmann, Professor Ruud Tromp, and Dr. Qiong Ma for helpful discussions; a manuscript reviewer for the suggestion of additional iDPC measurements; and Prof. Jeehwan Kim and Prof. Silvija Gradecak for equipment access. This research was primarily conducted on the traditional, unceded territory of the Wampanoag Nation. We acknowledge the painful history of forced removal from this territory, and we respect the many diverse indigenous people connected to this land.

## Author Contributions

K.R., G.V. and F.M.R. conceived the project. K.R. and J.D.T developed the epitaxial deposition and fabricated samples. G.V. extended the geometric convolution model in 2D/3D systems. K.R. and G.V. performed geometric convolution analysis and analysed the data. K.R. and J.D.T. performed SAED and bright field TEM imaging. A.M.B. performed 60 keV keV HRTEM imaging. A.K. and J.M.L performed 4D STEM imaging and STEM multislice simulations. K.R. performed 4D STEM data analysis, iDPC imaging, and multislice calculations with input from A.K. and T.P. J.D.T. performed AFM measurements. G.V. performed the density functional theory calculations and made atomic models, with input from P.N. and P.A. The manuscript was written with contributions from all authors.

## Competing Interests

The authors declare no competing interests.